\documentclass[12pt,english,floatfix,superscriptaddress,aps,prd,preprint,showkeys,nofootinbib]{revtex4}
\usepackage{amsmath}
\usepackage{amssymb}
\usepackage{amsbsy}
\usepackage{amsfonts}
\usepackage{amsopn}
\usepackage{amstext}
\usepackage{graphicx}
\usepackage[english]{babel}
\usepackage{color}
\usepackage{slashed}
\usepackage{esint}
\usepackage[dvips]{epsfig}
\usepackage[dvips]{graphicx}
\usepackage{float}
\usepackage{units}
\usepackage{textcomp}
\usepackage{wasysym}
\usepackage{hyperref}
\usepackage{slashed}
\begin{document}

\title{Scattering, absorption and greybody factor of scalar particles by Lorentz-violating charged black holes}

\author{F. M. Belchior}
\email{belchior@fisica.ufc.br}
\affiliation{Departamento de Física, Universidade Federal do Ceará (UFC),\\ Campus do Pici, 60455-760, Fortaleza - CE, Brazil.}

\begin{abstract}
In this work, we investigate the scattering and absorption of spin 0 particles for electrically charged black holes in two gravity models with spontaneous Lorentz symmetry breaking. The first one is the so-called bumblebee model that involves a vector field with a nonvanishing vacuum expectation value (VEV), while the second one involves a self-interacting Kalb-Ramond field coupled to gravity. For our purpose, we employ the partial waves method to compute the scattering cross-section and the absorption for these charged black holes. Moreover, we calculate the greybody factors (GFs) for spin 0 particles, showing the influence of both the LV parameter and electric charge.

\end{abstract}

\maketitle

%%%%%%%%%%%%%%%%%%%%%%%%%%%%%%%%%%%%%%%%%%%%%%%%%%%
\section{Introduction}
Recently, there has been a growing ambition to explore alternative gravity models to address unsolved issues from general relativity (GR). Among these issues, the lack of quantum gravity theory appears as the most challenging one, along with the accelerated expansion of the universe and other cosmological issues \cite{Sotiriou:2008rp, Clifton:2011jh, Akbar:2006mq}. Generally, modified gravity models are conceived by introducing more general geometric invariants into Einstein-Hilbert action or additional fields, such as the Brans-Dicke theory. In addition to Ricci scalar, one can add squared Ricci tensor $R_{\mu\nu}R^{\mu\nu}$, squared Riemann tensor $R_{\mu\nu\alpha\beta}R^{\mu\nu\alpha\beta}$ or the Gauss-Bonnet invariant $G=R^2-4R_{\mu\nu}R^{\mu\nu}+R_{\mu\nu\alpha\beta}R^{\mu\nu\alpha\beta}$ \cite{Kanti:1995vq, Nojiri:2005vv, Belchior:2024bcn}.

Investigations concerning exotic objects such as black holes and wormholes have become a fruitful line of research. In this context, one of the most notable predictions of general relativity (GR) is the existence of black holes (BHs) that represent objects from which nothing, not even light, can escape once the event horizon is crossed. The relevance of black holes goes beyond astrophysical interests, addressing pivotal issues in theoretical physics such as the nature of quantum gravity \cite{Hawking:1975vcx, Gibbons:1977mu, Bekenstein:1973ur}. As well-established in the literature, the quantum aspects of spacetime play an essential role in regions where the gravitational field is extremely strong like in the neighborhood of a black hole.

Another point of interest in black hole physics is due to recent observations of gravitational waves (GW). It is worth recalling that the first detection of these waves was due to the merger of a binary black hole system \cite{LIGOScientific:2016aoc}. Since then, there have been growing efforts to observe GW with increasing precision. Among these efforts, one highlights the LIGO and Virgo collaborations \cite{LIGOScientific:2017ycc, LIGOScientific:2020zkf, LIGOScientific:2018mvr, LIGOScientific:2016dsl, VIRGO:2014yos}.

Recently, the so-called bumblebee model has received considerable notoriety in the literature due to its simplicity and compatibility with curved space-time \cite{Bluhm:2004ep, Bailey:2006fd, Bluhm:2007bd, Kostelecky:2010ze}. It is a consistent way of implementing the effects of Lorentz symmetry breaking (LSB). Its construction is based on the presence of a vector field that assumes a nonzero vacuum expectation value (VEV), leading to spontaneous LSB. In the gravitational scenario, the action that describes such a model reads \cite{Maluf:2020kgf, Casana:2017jkc}
\begin{align}\label{action1}
S=\int d^4x\sqrt{-g}\Big[\frac{1}{2\kappa}(R+\epsilon B^\mu B^\nu R_{\mu\nu}) -\frac{1}{4}B_{\mu\nu}B^{\mu\nu}-V(B_\mu B^\mu-b^2)+\mathcal{L}_m\Big].
\end{align}
For this gravity model, some solutions of physical interest were investigated in recent works. For instance, the vacuum solution for a spherically symmetric metric was obtained in \cite{Maluf:2020kgf, Casana:2017jkc}. In \cite{Gullu:2020qzu} it was studied a black hole solution with global monopole, while in \cite{Liu:2024axg}, the authors reached a charged black solution by considering an extra coupling between the bumblebee and the $U(1)$ gauge fields through the following Lagrangian
\begin{align}
  \mathcal{L}_m=-\frac{1}{4} F_{\mu\nu}F^{\mu\nu}-\frac{\rho}{4} B^{\mu}B_{\mu} F_{\alpha\beta}F^{\alpha\beta}.
\end{align}
In addition to bumblebee model, another framework involves an antisymmetrical tensor known as Kalb-Ramond \cite{Altschul:2009ae, Maluf:2018jwc, Aashish:2019ykb}. Similar to bumblebee model, one assume a gravitational action as follows
\begin{align}\label{action2}
    S=\int d^4x\sqrt{-g}\bigg[\frac{1}{2\kappa}\bigg(R+\varepsilon\, B^{\mu\lambda}B^\nu\, _\lambda R_{\mu\nu}\bigg)-\frac{1}{12}H_{\lambda\mu\nu}H^{\lambda\mu\nu}-V(B_{\mu\nu}B^{\mu\nu}\pm b^2)+\mathcal{L}_m\bigg].
\end{align}
Some works were proposed to address the physical aspects of Kalb-Ramond gravity. In this conjecture, one can cite the vacuum solution for a spherically symmetric metric derived in \cite{Lessa:2019bgi, Yang:2023wtu}, while a charged black hole geometry was obtained, and a black hole solution with global monopole was obtained in \cite{Belchior:2025xam}. Particularly, to derive a charged black hole solution, it is necessary to add a nonminimal coupling between the $U(1)$ gauge and the Kalb-Ramond fields \cite{Duan:2023gng}. Then, the matter Lagrangian for this case reads
\begin{align}
  \mathcal{L}_m=-\frac{1}{4} F_{\mu\nu}F^{\mu\nu}-\frac{\chi}{4} B^{\mu\nu}B^{\alpha\beta} F_{\mu\nu}F_{\alpha\beta}.
\end{align}

In this thematic, inspired by the aforementioned works in which charged black holes solutions were obtained in Lorentz-breaking gravities, we seek to investigate scattering, absorption and greybody factor of scalar particles. Fou our research, we will make use of the partial wave method, which it has been employed in recent works \cite{Sanchez:1976xm, Crispino:2009ki,Anacleto:2020zhp, Anacleto:2019tdj, Pitelli:2017bgx, Anacleto:2017kmg, Anacleto:2022shk}. Besides, we will follows some recent papers \cite{Boonserm:2008zg, Okyay:2021nnh, Guo:2023nkd, Wu:2024ldo, Sucu:2025lqa, Li:2024xyu, Campos:2023zmg, al-Badawi:2024pdx, Richarte:2021fbi} to investigate greybody of massless scalar field.

The study of wave scattering and absorption by black holes plays a central role in our understanding of quantum and classical aspects of gravity.  In general relativity, the interaction of scalar, electromagnetic, or gravitational waves with black-hole geometries encodes key information about spacetime curvature, causal structure, and thermodynamic properties. When Lorentz symmetry is violated, these processes become even more significant, as wave propagation directly probes how fundamental symmetries shape the dynamics of fields in curved backgrounds.

The paper is outlined as follows: in section \ref{s2}, we present the metrics that describe the black hole with electric charged in LSB gravities. In section \ref{s3}, one derives the equation of motion of a probe massless scalar field for a general spherically symmetric space-time. In section \ref{s4}, one examines the scattering and absorption by utilizing the partial wave method. The greybody factors are calculated in section \ref{s5}. Finally, our final considerations and future perspectives are presented in section \ref{s6}

\section{Charged black hole solutions}\label{s2}
In this section, we introduce the charged black hole with solutions that arise from Lorentz-violating models for the purpose of determining the differential scattering cross section and absorption in the next section.

\subsection{Solution 1: charged black hole in bumblebee gravity}
Initially, one assumes following the static and spherically symmetrical line element \cite{Liu:2024axg}
\begin{align}\label{metric1}
    ds^2=-A(r)dt^2+B(r)dr^2+r^2(d\theta^2+\sin^2{\theta}d\phi^2),
\end{align}
where the functions $A(r)$ and $B(r)$ read
\begin{align}\label{sol1a}
    A(r)=1-\frac{2M}{r}+\frac{2(1+l)\,Q^2}{2r^2\,(2+l)},
\end{align}
and
\begin{align}\label{sol1b}
    B(r)=\frac{1+l}{A(r)},
\end{align}
Herein, we have defined the LV parameter $l$, which is related to the bumblebee VEV. To obtain the aforementioned geometry, the authors considered a radial configuration for bumblebee field, i.e., $B_\mu(r)=(0,b,0,0)$. In this case, the bumblebee field configuration is frozen to its VEV.  As we can see, the above solution is similar to the Reissner–Nordström (RN) solution, recovering it when $l=0$. Besides, there are also horizons, namely
\begin{align}
    r_{\pm}=M\pm\sqrt{M^2-\frac{2(1+l)}{2+l}\,Q^2}
\end{align}
where $r_{-}$ and $r_{+}$ represent the event and the Cauchy horizons, respectively. It is necessary that the mass and charge parameter of the black hole satisfy the following relation
\begin{align}
    \frac{Q^2}{M^2}\leq \frac{2(1+l)}{2+l}
\end{align}

\subsection{Solution 2: charged black hole in Kalb-Ramond gravity}
For the solution 2, one considers the following the static and spherically symmetrical line black hole \cite{Duan:2023gng}
\begin{align}\label{metric2}
    ds^2=-A(r)dt^2+\frac{dr^2}{A(r)}+r^2(d\theta^2+\sin^2{\theta}d\phi^2),
\end{align}
where the function $A(r)$ is given by
\begin{align}\label{sol2}
    A(r)=\frac{1}{1-\gamma}-\frac{2M}{r}+\frac{Q^2}{r^2\,(1-\gamma)^2},
\end{align}
Similar to the previous case, the LV parameter $\gamma$ is related to Kalb-Ramond VEV. This solution was found by considering a pseudo-electric configuration in which the Kalb-Ramond field $B_{\mu\nu}$ is frozen to its VEV $b_{\mu\nu}$, so that one writes its explicit form as follows \cite{Duan:2023gng}:
\begin{align}\label{KRconfig}
b_{\mu\nu}=b_{01}=-b_{10}=\frac{\vert b \vert}{\sqrt{2}}.    
\end{align}
This configuration leads to the constant norm $b_{\mu\nu}b^{\mu\nu}=\vert b \vert^2$. Again, we recover the Reissner–Nordström (RN) solution when taking $\gamma=0$ in (\ref{sol2}). The event and the Cauchy horizons read
\begin{align}
    r_{\pm}=M(1-\gamma)\pm\sqrt{M^2(1-\gamma)^2-\frac{Q^2}{1-\gamma}}
\end{align}
In this case, one has the following relation 
\begin{align}
    \frac{Q^2}{M^2}\leq (1-\gamma)^3
\end{align}
Having presented, we will analyze for a massless scalar field in a generic spherically symmetric spacetime and posteriorly analyze scattering, absorption and greybody factors.

\section{Probe real scalar field in spherically symmetric spacetime}\label{s3}

Once we have defined the black hole solutions in LV gravities, we are able to look into the behavior of a scalar field in these geometries. Specifically, we are interested in determining the differential scattering cross-section for black hole solutions presented previously by means of the partial wave method in the low-frequency regime (see f.e.\cite{Anacleto:2020zhp, Anacleto:2019tdj, Pitelli:2017bgx, Anacleto:2017kmg, Anacleto:2022shk}). To do so, one assumes the following action for a minimally coupled non-massive scalar field to gravity
\begin{align}\label{sac}
    S_{CE}=-\frac{1}{2}\int d^4x\sqrt{-g}\nabla_\mu\psi\nabla^\mu\psi.
\end{align}
We stand out that the field $\psi$ is a probe field, so that any back-reaction can be neglected. Thereby, by varying the action (\ref{sac}) with respect to the field $\psi$, we obtain the Klein-Gordon equation in curved space-time
\begin{align}\label{ce}
    \nabla_\mu\nabla^\mu\psi=\partial_\mu (\sqrt{-g} g^{\mu\nu}\partial_\nu \psi)=0.
\end{align}
Now, let us consider a generic static and spherically symmetric metric given by
\begin{align}\label{metric3}
    ds^2=-A(r)dt^2+B(r)dr^2+r^2(d\theta^2+\sin^2{\theta}d\phi^2).
\end{align}
For this metric, one has $\sqrt{-g}=A^{1/2}\,B^{1/2}\,r^2\sin\theta$, thus (\ref{ce}) becomes
\begin{align}\label{eqce}
    -A^{-1/2}\,B^{1/2}\,r^2\,\frac{\partial^2\psi}{\partial t^2}+\frac{\partial}{\partial r}\bigg(A^{1/2}\,B^{-1/2}\,r^2  \frac{\partial\psi}{\partial r}\bigg)\nonumber\\+A^{1/2}\,B^{1/2}\bigg[\frac{1}{\sin\theta}\frac{\partial}{\partial \theta}\bigg(\sin\theta\frac{\partial\psi}{\partial \theta}\bigg)+\frac{1}{\sin\theta}\frac{\partial^2\psi}{\partial \phi^2}\bigg]=0.
\end{align}
We can employ the following separation of variables
\begin{align}\label{sv}
    \psi(t,r,\theta,\phi)=\frac{\mathcal{R}(r)}{r}Y_{lm}(\theta,\phi)e^{-i\omega t},
\end{align}
where $w$ is the frequency and $Y_{lm}(\theta,\phi)$ represent the spherical harmonics. With (\ref{sv}), the radial equation looks like
\begin{align}\label{req}
    \mathcal{R}''(r)
+\frac{1}{2}\left(\frac{A'(r)}{A(r)}-\frac{B'(r)}{B(r)}\right)\mathcal{R}'(r)
+\frac{B(r)}{A(r)}\left[\omega^2-V_{eff}(r)\right]\mathcal{R}(r)=0
\end{align}
where $V_{eff}$ stands for the effective potential
\begin{align}
 V_{eff}(r)=A(r)\frac{\lambda(\lambda+1)}{r^2}+\frac{1}{2r}\frac{d}{dr}\!\left(\frac{A(r)}{B(r)}\right),  
\end{align}
where $\lambda$ represents the angular momentum quantum number. With the radial equation as well as the effective potential in hands, the next step is to delve into the process of scattering and absorption for the two black hole solutions discussed in the previous section by means of the partial wave method.

\section{Scattering and absorption}\label{s4}

Let us delve into the analysis of scattering and absorption of massless scalar particles. The scattering of scalar waves is related to the response of a black hole to external perturbations, which can have origin from scalar, vector or spinor fields. In this context, the scattering amplitude and differential cross section depend sensitively on the form of the effective potential surrounding the horizon. On the other hand, the absorption cross section measures the probability that an incoming wave is captured by the black hole rather than scattered away. At the quantum level, this quantity determines the fraction of Hawking radiation emitted at infinity after transmission through the effective potential barrier that surrounds the horizon.  

In the present section, we will investigate these processes in charged black holes in two distinct frameworks, bumblebee gravity and Kalb-Ramond (KR) gravity presented in the section \ref{s2}. As we highlighted, these models provide complementary examples of how vector and antisymmetric tensor condensates affect the spacetime curvature and, consequently, the scattering pattern of scalar waves. In Lorentz-violating spacetimes, the deformation of the metric function $F(r)$ alters both the location of the potential peak and the surface gravity, leading to characteristic changes in absorption cross section ($\sigma(\omega)$)  and in the Hawking temperature $T_H$. By examining these quantities, one gains insight into how Lorentz-violating backgrounds influence energy emission and information transfer from the horizon.

\subsection{Solution 1}
For the first solution, we use the relation $B(r)=\frac{1+l}{A(r)}$, so that the Eq. (\ref{req}) becomes
\begin{align}\label{ece1}
    A\frac{d}{dr}\bigg(A\frac{d\mathcal{R}}{dr}\bigg)+(1+\gamma)[\omega^2-\mathcal{V}_{eff}(r)]\mathcal{R}=0,
\end{align}
where $V_{eff}$ is the effective potential, namely
\begin{align}
  V_{eff}(r)=\frac{A}{r(1+l)}\frac{dA}{dr} + \frac{A\  \lambda(\lambda+1)}{r^2}. 
\end{align}

At this point, it is convenient to apply the transformation $\chi(r)=\sqrt{A(r)}\mathcal{R}(r)$, so that Eq. (\ref{ece1}) turns into a Schrödinger-like equation. Thus, one has
\begin{align}\label{echi1}
    \frac{d^2\chi(r)}{dr^2}+U(r)\chi(r)=0,
\end{align}
where
\begin{align}
    U(r)=\frac{A^{\prime 2}}{4A^2}-\frac{A^{\prime \prime}}{2A}+\frac{(1+\gamma)(\omega^2-\mathcal{V}_{eff})}{A^2}.
\end{align}
Thus, by using the solution (\ref{sol1a}), we arrive at
\begin{align}\label{U1}
U(r)&=(1+l)\, \omega ^2+\frac{4 (1+l) M \omega ^2}{r}+\frac{12\ell^2}{r^2}\nonumber\\&+\frac{2 M \omega ^2 \left(16 (1+l) M^2-3 (l +2) Q^2\right)-2 (1+l)\, \lambda (\lambda+1) M}{r^3}+\cdots,
\end{align}
with
\begin{align}
    \ell^2=\bigg[(1+l) M^2-\frac{(2+l)Q^2}{12}\bigg]\omega ^2-\frac{\lambda(\lambda+1)(1+l)}{12}.
\end{align}

In (\ref{U1}), we have performed a power series expansion in $1/r$,
defining $\ell^2$ as being the change in the coefficient of $1/r^2$ (including only contributions involving the quantities $\ell$ and $\omega$). One notices that as $r$ tends to infinity, the potential $U(r)$ tends to zero, so that the asymptotic behavior is satisfied. If the phase shift is known, it is straightforward to obtain the scattering amplitude by means of the following partial wave representation.
\begin{align}
    f(\theta)=\frac{1}{2i\omega}\sum_{\lambda=o}^\infty (2\lambda+1)(e^{2i\delta_\lambda}-1)P_\lambda(\cos\theta),
\end{align}
while we compute the differential scattering cross section through the expression
\begin{align}
  \frac{d\sigma}{d\theta}=\vert f(\theta)\vert^2.  
\end{align}
Additionally, we can obtain the phase shift by utilizing
\begin{align}
    \delta_l=2(\lambda-\ell).
\end{align}
Thus, in the limit $\lambda \rightarrow 0$, we find
\begin{align}\label{d1}
 \delta_0=-2\sqrt{(1+l) M^2-\frac{(2+l)Q^2}{12}}\,\omega.   
\end{align}
It is simple to observe that when we take the limit $\ell \rightarrow 0$, the phase shifts tend towards nonzero terms, naturally leading to a correct result for the differential cross-section in the small-angle limit. One should point out that the representation is poorly convergent what makes the task of performing m the sum of the series pretty difficult. Hence, we have a problem, where an infinite number of Legendre polynomials are required to obtain divergences in $\theta=0$. In order to bypass this problem, it is utilized a reduced series, which is less divergent in $\theta=0$. Thus, one writes explicitly
\begin{align}
(1-\cos\theta)^m\,f(\theta)=\sum_{m} a_m\, P_m(\cos\theta).  
\end{align}
Additionally, we determine the differential scattering cross-section by applying the following equation 
\begin{align}
\frac{d\sigma}{d\Omega}=\frac{1}{2\pi \omega}  \sum_{\lambda} (2m+1)(e^{2i\delta_\lambda}-1)\frac{P_\lambda(\cos\theta)}{1-cos\theta}.  
\end{align}
Therefore, by considering (\ref{d1}), we obtain the cross-section for $m$ ($m=0,1$) as follows
\begin{align}
    \frac{d\sigma}{d\Omega}=\frac{4\delta_0^2}{\omega^2\theta^4}=\frac{16}{\theta^4}\bigg[(1+l) M^2-\frac{(2+l)Q^2}{12}\bigg]
\end{align}
As we can verify, the differential scattering cross-section is increased by the LV effects, while it is decreased by the electrical charge. On the other hand, let us employ an alternative path to obtain the same phase shift. As a starting point, one considers the Born approximation formula, namely
\begin{align}
    \delta_l=\frac{\omega}{2}\int_0^\infty dr r^2 J_l^2(\omega r) u(r).
\end{align}
where $J_l(\omega r)$ stand for the spherical Bessel functions of the first kind and $u(r)$ is the potential given by
\begin{align}
u(r)=\frac{2 M \omega ^2 \left(16 (1+l) M^2-3 (l +2) Q^2\right)-2 (1+l)\, \lambda (\lambda+1) M}{r^3}+\cdots.    
\end{align}
Now, let turn our attention to the phenomena of absorption ahead. Let us compute the absorption cross section for solution 1 at the low
frequency limit. In this case, we write from quantum mechanic
\begin{align}
    \sigma_{abs}&=\frac{\pi}{\omega^2}\sum_{\lambda=0}^{\infty}(2\lambda+1)(1-e^{2i\delta_\lambda})=\frac{4\pi}{\omega^2}\sum_{\lambda=0}^{\infty}(2\lambda+1)\sin^2{\delta_\lambda}\nonumber\\&=\frac{4\pi}{\omega^2}\bigg[\sin^2{\delta_0}+\sum_{\lambda=1}^{\infty}(2\lambda+1)\sin^2{\delta_\lambda}\bigg]
\end{align}
When taking the limit $\omega\rightarrow 0$, we obtain that the absorption reads
\begin{align}
   \sigma_{abs}=\frac{4\pi\delta_0^2}{\omega^2}=16\pi\bigg[(1+l) M^2-\frac{(2+l)Q^2}{12}\bigg] 
\end{align}

As we can note, there is a contribution of both the LV parameter $l$ and of the electrical charge $Q$. The absorption decreases as we increase the parameters $l$ and $Q$. If we set $l=0$, we recover the result obtained for a charged black hole \cite{Crispino:2009ki}, and when $l=Q=0$, the result obtained for the Schwarzschild black hole is obtained \cite{Sanchez:1976xm}.

\subsection{Solution 2}
For the second solution, we have the relation $B(r)=\frac{1}{A(r)}$ and the Eq. (\ref{req}) reads
\begin{align}\label{ece2}
    A\frac{d}{dr}\bigg(A\frac{d\mathcal{R}}{dr}\bigg)+[\omega^2-\mathcal{V}_{eff}(r)]\mathcal{R}=0,
\end{align}
where $\mathcal{V}_{eff}$ is the effective potential defined as
\begin{align}
  \mathcal{V}_{eff}(r)=\frac{A}{r}\frac{dA}{dr} + \frac{A\  \lambda(\lambda+1)}{r^2}. 
\end{align}

To proceed further, let write Eq.(\ref{ece2}) into a Schrödinger-like form using again the transformation $\chi(r)=\sqrt{A(r)}\mathcal{R}(r)$, so that we have
\begin{align}\label{echi}
    \frac{d^2\chi(r)}{dr^2}+\mathcal{U}(r)\chi(r)=0,
\end{align}
where
\begin{align}
    \mathcal{U}(r)=\frac{A^{\prime 2}}{4A^2}-\frac{A^{\prime \prime}}{2A}+\frac{(\omega^2-\mathcal{V}_{eff})}{A^2}.
\end{align}

One can perform power series in $1/r$ in (\ref{echi}), so that we have
\begin{align}
\mathcal{U}(r)&=(1-\gamma)^2 \omega ^2+\frac{4 (1-\gamma)^3 M \omega ^2}{r }+\frac{12\ell^2}{r^2}\nonumber\\&-\frac{2 (1-\gamma)^2 M \left(\lambda(\lambda+1)+2 \omega ^2 \left(3 Q^2-8 (1-\gamma)^3 M^2\right)\right)}{r^3}+\cdots,
\end{align}
with
\begin{align}
    \ell^2=(1-\gamma)\bigg[(1-\gamma)^3 M^2-\frac{Q^2}{6}\bigg]\omega^2-\frac{\lambda(\lambda+1)(1-\gamma)}{12}.
\end{align}
In this case, we can use the previous definitions and write the differential cross section as follows
\begin{align}
    \frac{d\sigma}{d\Omega}=\frac{4\delta_0^2}{\omega^2\theta^4}=\frac{16(1-\gamma)}{\theta^4}\bigg[(1-\gamma)^3 M^2-\frac{Q^2}{6}\bigg]
\end{align}
From the above result, let us observe that the differential scattering cross-section decreases as the LV parameter and the electrical charge grows. Additionally, the absorption reads
\begin{align}
   \sigma_{abs}=\frac{4\pi\delta_0^2}{\omega^2}=16\pi\,(1-\gamma)\bigg[(1-\gamma)^3 M^2-\frac{Q^2}{6}\bigg] 
\end{align}
The absorption exhibits the same behavior as the scattering cross-section, decreasing as we increase $\gamma$ and $Q$.

\section{Greybody factor}\label{s5}

An important result shown by Hawking is that black holes are able to emit radiation, which is known as Hawking radiation. Such an effect is the consequence of quantum effects that fields suffer in the neighborhood of the event horizon. After being emitted by a black hole, this radiation interacts with the curved spacetime around it, which acts as a gravitational barrier. As a consequence, this interaction alters the spectrum of the radiation, so that an observer spotted at an infinite distance detects a deviation from the original Hawking radiation. In this context, the greybody factor (GF) appears as the quantity that measures this deviation. The greybody factor links the microscopic dynamics of wave propagation to the macroscopic thermodynamics of black holes.

Therefore, in this section, we are interested in obtaining the GF of the scalar field for the two black hole solutions discussed so far. Our departure point is the transmission probability defined as follows \cite{Boonserm:2008zg, Okyay:2021nnh, Guo:2023nkd, Wu:2024ldo}
\begin{align}
 \sigma(\omega)=\mathrm{sech}^2\bigg(\int_{-\infty}^{\infty}\rho(r_{\ast}) dr_{\ast}\bigg)   \end{align}
where $dr_{\ast}$ is the tortoise coordinate and
\begin{align}
\rho(r_{\ast})=\frac{1}{2h}\sqrt{\bigg(\frac{d h(r_{\ast})}{dr_{\ast}}\bigg)^2+(\omega^2-V_{eff}(r)-h^2(r_{\ast}))^2}    
\end{align}
In the above formula, $h(r_{\ast})$ represents a positive function satisfying $h(\infty)=h(-\infty)=w$
\begin{align}
\sigma(\omega)=\mathrm{sech}^2\bigg(\frac{1}{2\omega}\int_{r_H}^{\infty}V_{eff}(r_{\ast})\, dr_{\ast} \bigg)   
\end{align}
It is simple to observe that the radial functions plays a relevant role in establishing the relationship between the greybody factor and the effective potential.
\subsection{Solution 1}
Initially, let us rewrite Eq.(\ref{ece1}) as follows
\begin{align}\label{ece11}
    \frac{d^2\mathcal{R}}{dr^{\ast 2}}+[\omega^2-\mathcal{V}_{eff}(r)]\mathcal{R}=0,
\end{align}
where we have defined the tortoise coordinate $\frac{dr^{\ast}}{dr}=\frac{1+l}{A}$ and $V_{eff}$ is the effective potential becomes
\begin{align}\label{eff11}
  V_{eff}(r)=\frac{A}{r(1+l)}\frac{dA}{dr} + \frac{A\  \lambda(\lambda+1)}{r^2}. 
\end{align}

Thus, the GF with the potential effective (\ref{eff11}) is given by 
\begin{align}
  \sigma(\omega)=\mathrm{sech}^2\bigg[\frac{\sqrt{1+l}}{2\omega}\int_{r_{-}}^{r_{+}}\bigg(\frac{A^{\prime}}{r\,(1+l)}+\frac{\lambda(\lambda+1)}{r^2}\bigg) dr \bigg]  
\end{align}
Using the function \ref{sol1a}, we arrive at
\begin{align}
  T(\omega)&=\mathrm{sech}^2\bigg[\frac{\sqrt{1+l}}{2\omega} \,\Sigma \bigg]  
\end{align}
where
\begin{align}
  \Sigma=\frac{m (m+1)}{\sqrt{M^2-2 Q^2}+M}-\frac{4 Q^2}{3 (2+l) \left(\sqrt{M^2-2 Q^2}+M\right)^3}+\frac{M}{(1+l) \left(\sqrt{M^2-2 Q^2}+M\right)^2} 
\end{align}

\begin{figure}[ht!]
\begin{center}
\begin{tabular}{ccc}
\includegraphics[height=5cm]{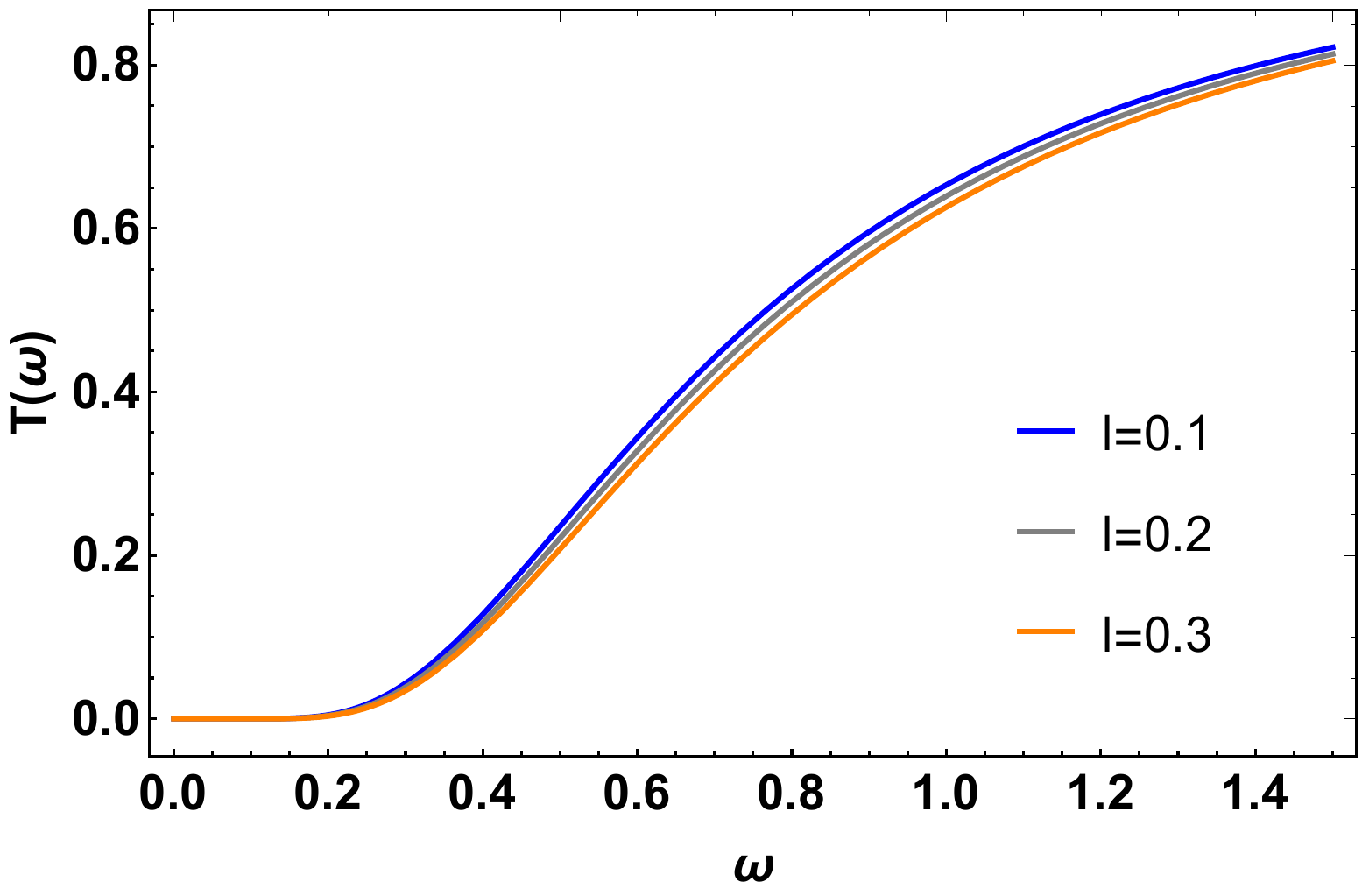} 
\includegraphics[height=5cm]{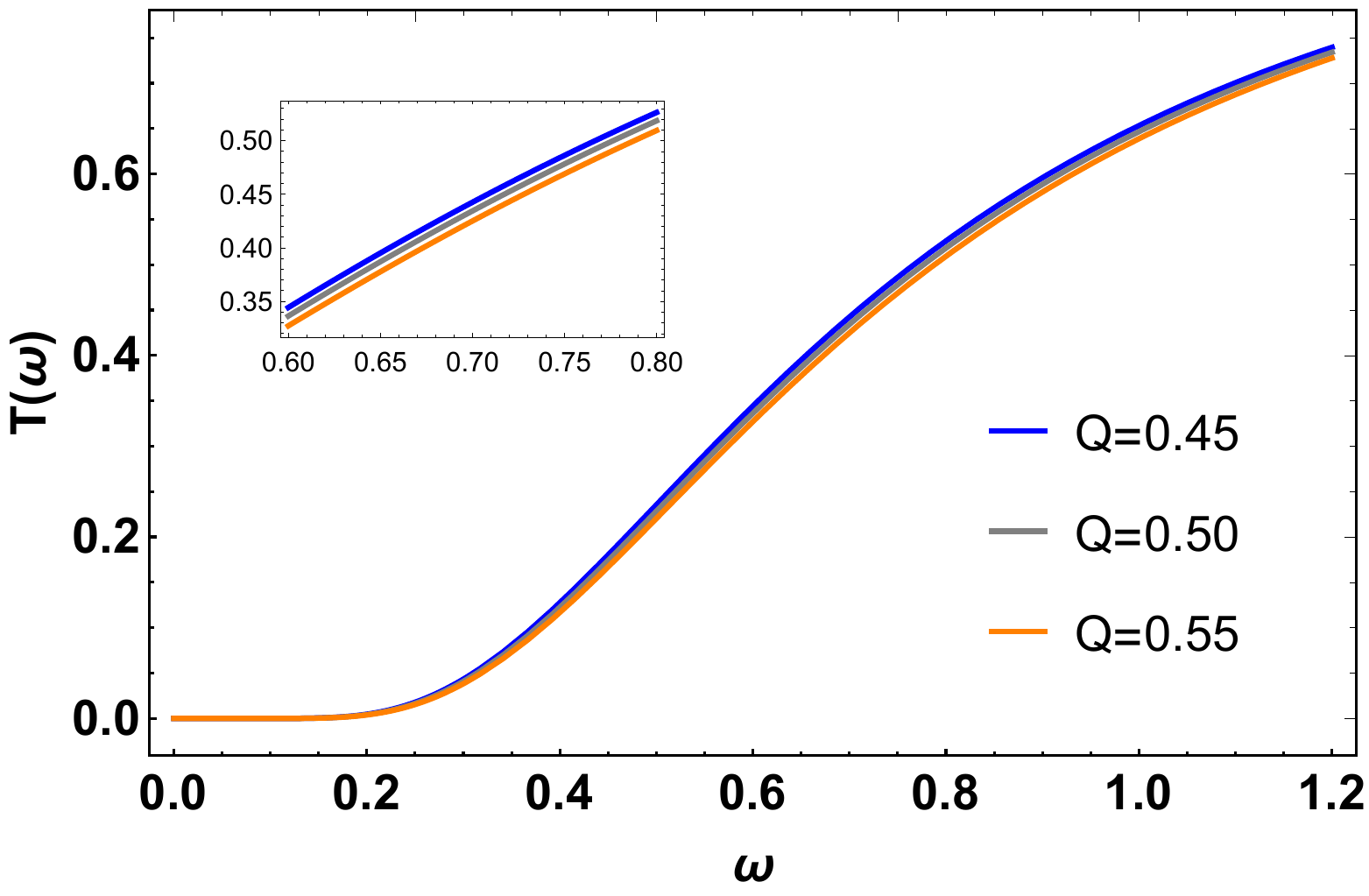}\\
\end{tabular}
\end{center}
\vspace{-0.5cm}
\caption{For solution 1. (a) Greybody factor for $\lambda=0$. (b) Greybody factor for $l=0.1$. 
\label{fig1}}
\end{figure}

One notices that the GF factor depends on both the LV parameter $l$ and on the electric charge $Q$. For a better analysis, the behavior of GF is illustrated in Fig.(\ref{fig1}). It is straightforward to observe that radiation is reflected totally and there is no transmission when the frequency is at a lower level. On the other hand, the tunneling effect occurs when the frequency is increased, allowing the radiation to pass through the potential barrier. Furthermore, the greybody factor decreases as the LV parameter and the electric assume larger values. Such a situation can be physically interpreted as follows: when we increase the parameters $l$ and $Q$, the height of the potential barrier also increases leading to a lower probability of transmission to the incident wave. 

\subsection{Solution 2}
For the solution 2, we utilize the the tortoise coordinate defined as $\frac{dr^{\ast}}{dr}=\frac{1}{A}$, so that the Eq.(\ref{ece2}) reduces to
\begin{align}\label{ece22}
    \frac{d^2\mathcal{R}}{dr^{\ast 2}}+[\omega^2-\mathcal{V}_{eff}(r)]\mathcal{R}=0,
\end{align}
where $\mathcal{V}_{eff}$ is the effective potential reads
\begin{align}\label{eff22}
  \mathcal{V}_{eff}(r)=\frac{A}{r}\frac{dA}{dr} + \frac{A\  \lambda(\lambda+1)}{r^2}. 
\end{align}

the GF with the potential effective (\ref{eff22}) is given by 
\begin{align}
  \sigma(\omega)=\mathrm{sech}^2\bigg[\frac{1}{2\omega}\int_{r_{-}}^{r_{+}}\bigg(\frac{A^{\prime}}{r}+\frac{\lambda(\lambda+1)}{r^2}\bigg) dr \bigg]  
\end{align}
Using the function \ref{sol2}, we arrive at
\begin{align}
  T(\omega)&=\mathrm{sech}^2\bigg[\frac{1}{2\omega} \,\Sigma \bigg]  
\end{align}
where
\begin{align}
  \Sigma=&-\frac{2 (\gamma -1) Q^2}{3 \left(\sqrt{(\gamma -1) \left((\gamma -1)^3 M^2+Q^2\right)}-(\gamma -1)^2 M\right)^3}\nonumber\\&+\frac{m (m+1) \left(\frac{\sqrt{(\gamma -1) \left((\gamma -1)^3 M^2+Q^2\right)}}{\gamma -1}-\gamma  M+M\right)+M}{\left(\frac{\sqrt{(\gamma -1) \left((\gamma -1)^3 M^2+Q^2\right)}}{\gamma -1}-\gamma  M+M\right)^2}
\end{align}

Similar to the previous case, we perceive that the GF factor depends on both the LV parameter $\gamma$ and on the electric charge $Q$. In this case, the behavior of GF is illustrated in Fig.(\ref{fig1}). Again, the tunneling effect only occurs when the frequency is increased. Like the previous case, the greybody factor decreases as both the LV parameter and the electric assumes assume larger values.

\begin{figure}[ht!]
\begin{center}
\begin{tabular}{ccc}
\includegraphics[height=5cm]{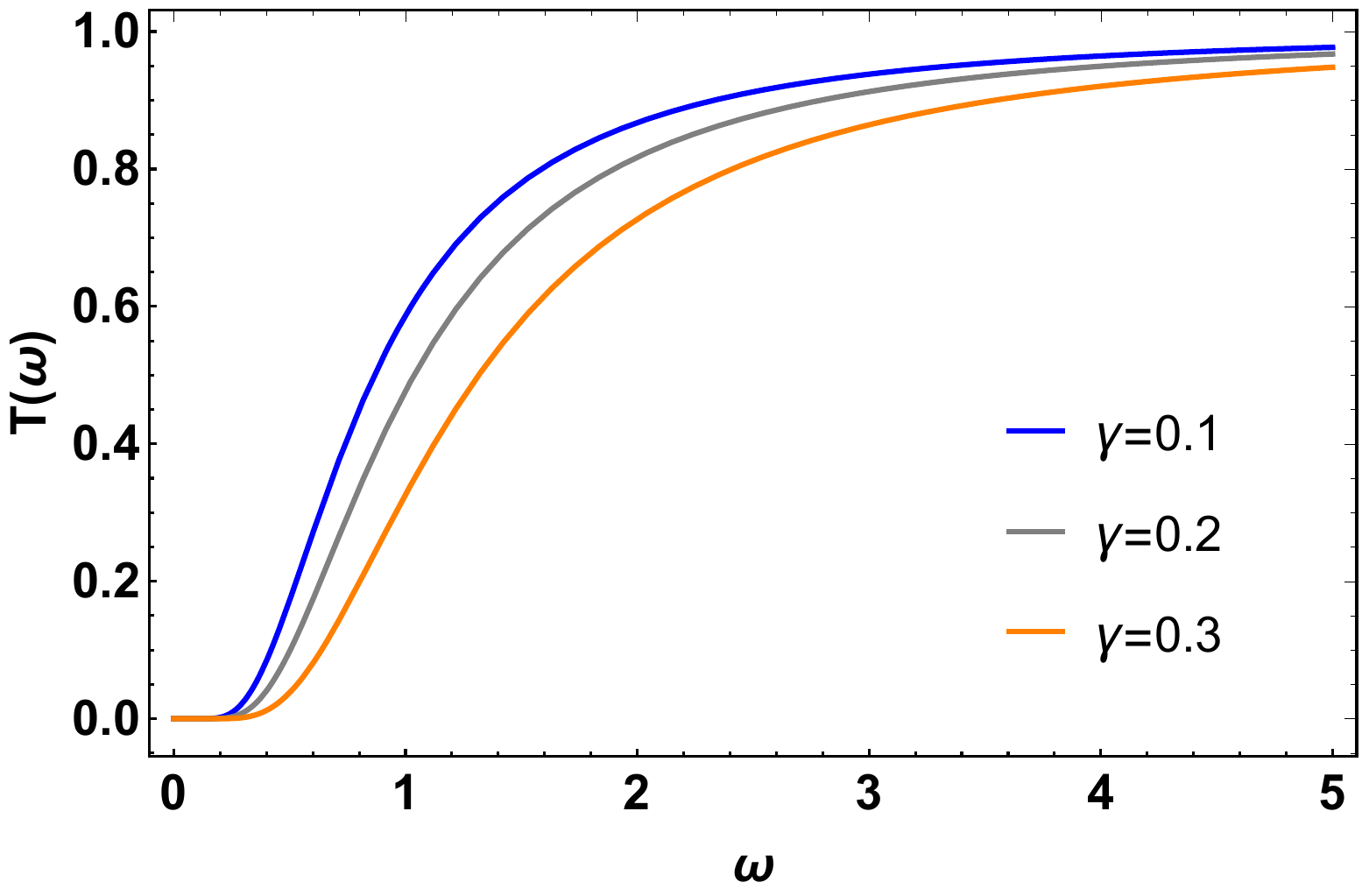} 
\includegraphics[height=5cm]{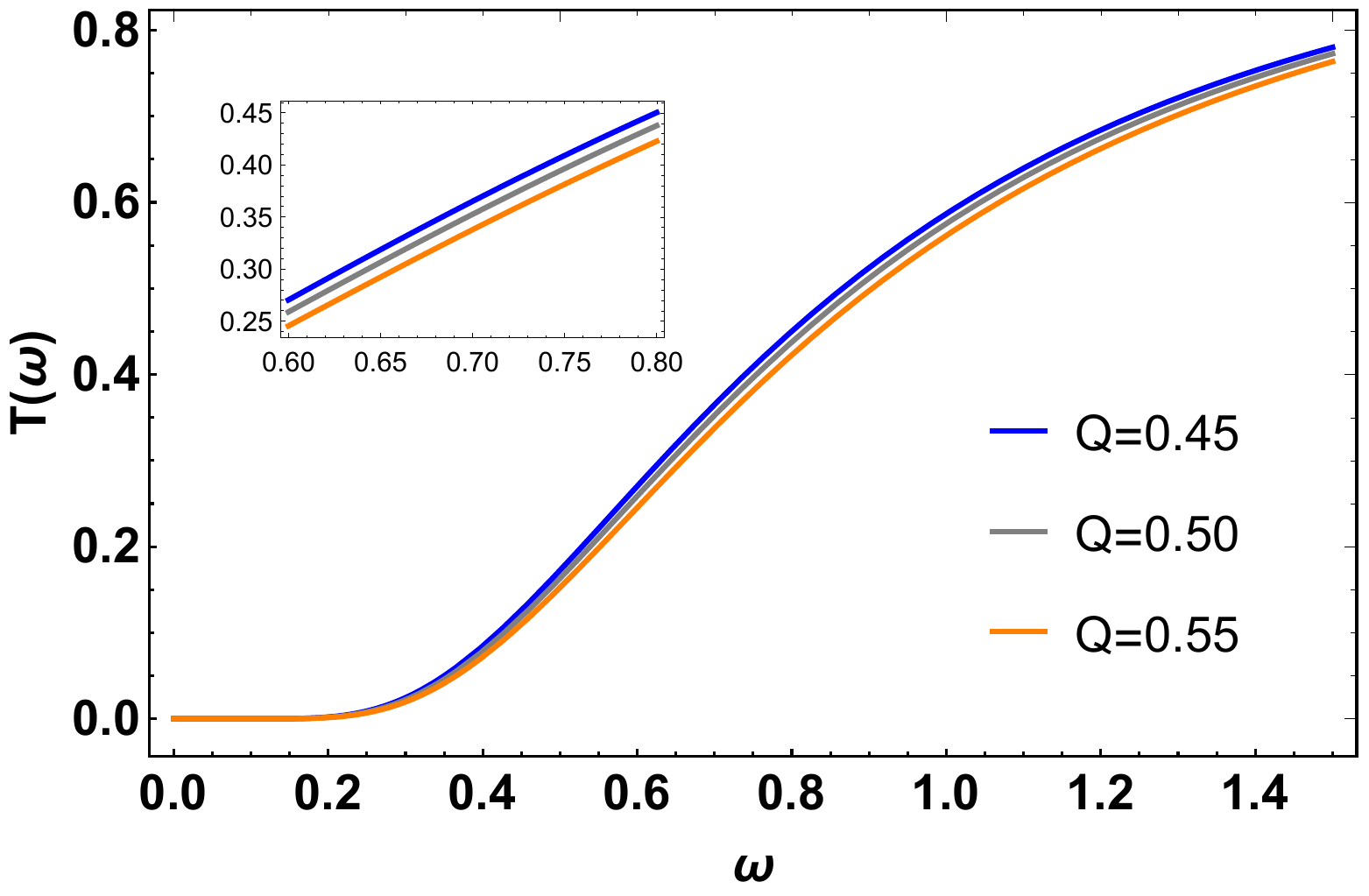}\\
(a)\hspace{6.7cm}(b)\\
\end{tabular}
\end{center}
\vspace{-0.5cm}
\caption{For solution 2. (a) Greybody factor for $\lambda=0$. (b) Greybody factor for $\gamma=0.1$.
\label{fig2}}
\end{figure}

Therefore, we performed a comparative analysis of the scattering, absorption, and greybody factors of a massless scalar field in two distinct LV charged black-hole solutions produced in the bumblebee and Kalb-Ramond gravity models. We show that with their similar origin in spontaneous LSV, these two theories display a similar physical trends. Both background fields modify the spacetime curvature. In the bumblebee model, the vector field induces an anisotropic modification of the radial curvature, which effectively stiffs the spacetime and impeding wave transmission.  Similarly, the LV Kalb-Ramond field acts as a "stiffing agent", increasing the curvature strength and hindering the propagation of scalar modes. Besides, the greybody factor decreases with electric charge \(Q\) for both solutions (Figs.1(b) and 2(b)). Such a trend is plausible for the parameters adopted. 

\begin{figure}[ht!]
\begin{center}
\begin{tabular}{ccc}
\includegraphics[height=5cm]{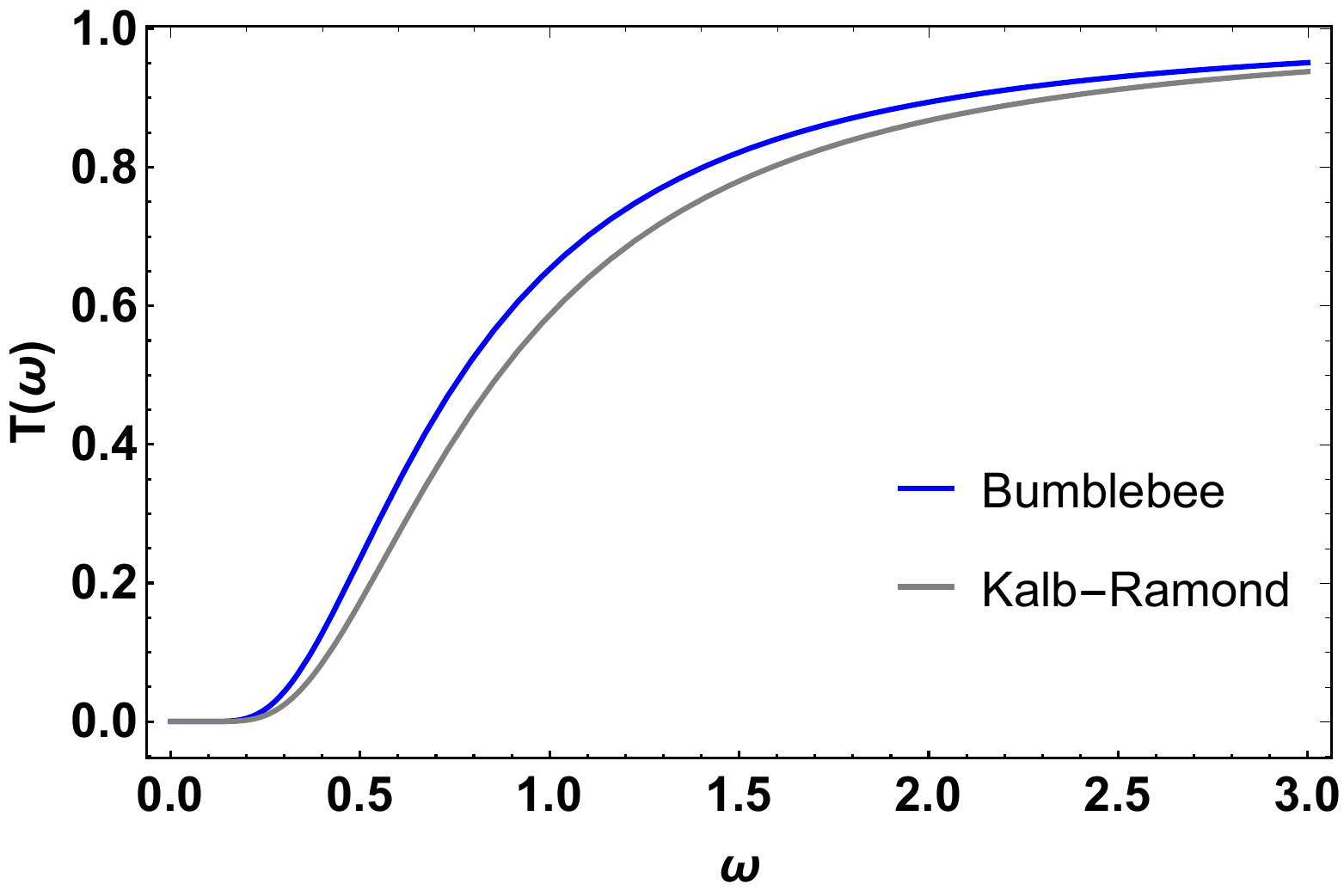} 
\end{tabular}
\end{center}
\vspace{-0.5cm}
\caption{ Comparison between the greybody factor for bumblebee model with $\lambda=1$ and $l=0.1$ and the greybody factor for Kalb-Ramond model for $\lambda=1$ and $\gamma=0.1$. 
\label{fig3}}
\end{figure}

Despite their similarities, the Bumblebee and Kalb-Ramond gravity models has a quantitative difference. We note in Fig.\ref{fig3} that the amplitude of the greybody factor for Bumblebee model is greater than for Kalb-Ramond model.

\begin{figure}[ht!]
\begin{center}
\begin{tabular}{ccc}
\includegraphics[height=5cm]{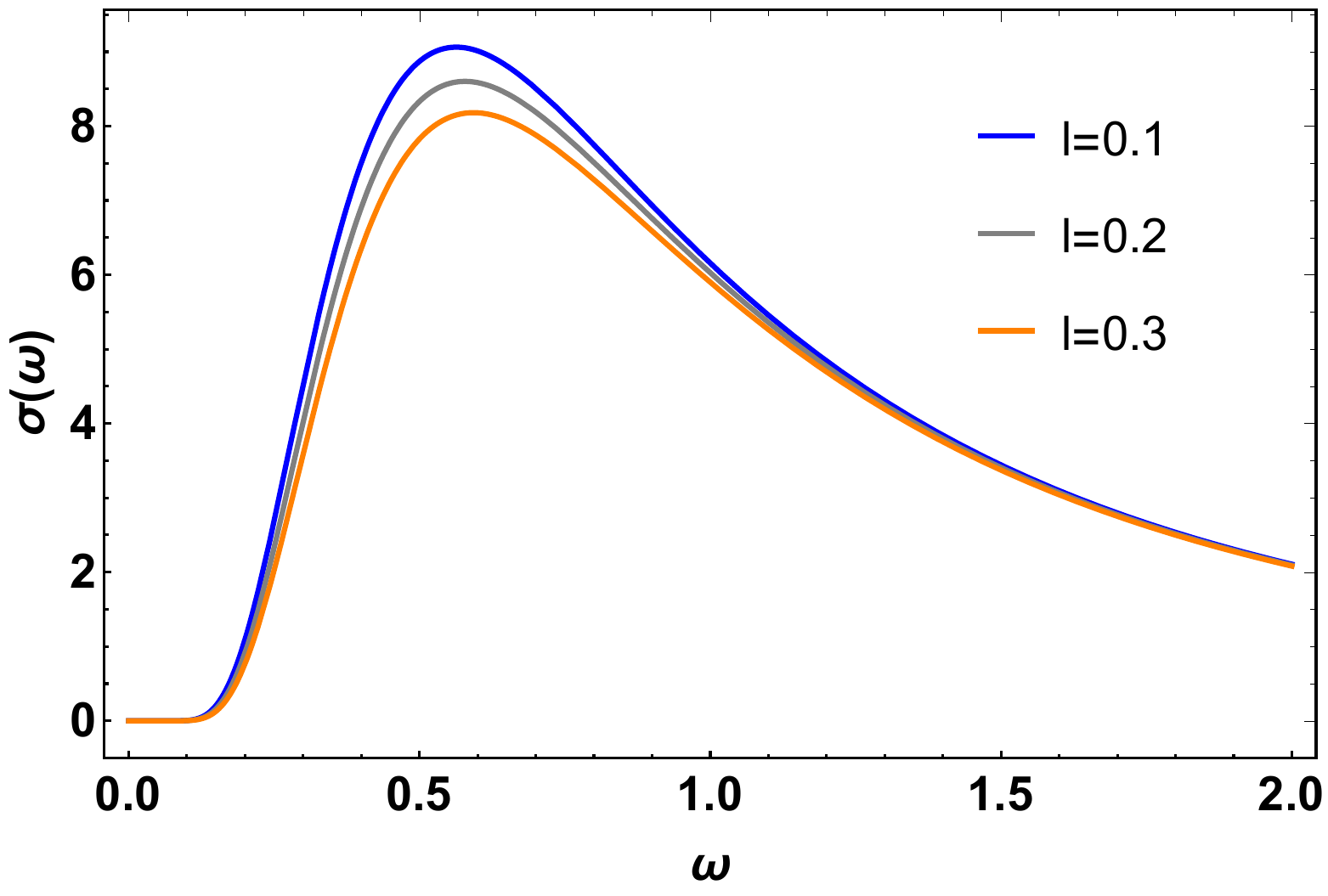} 
\includegraphics[height=5cm]{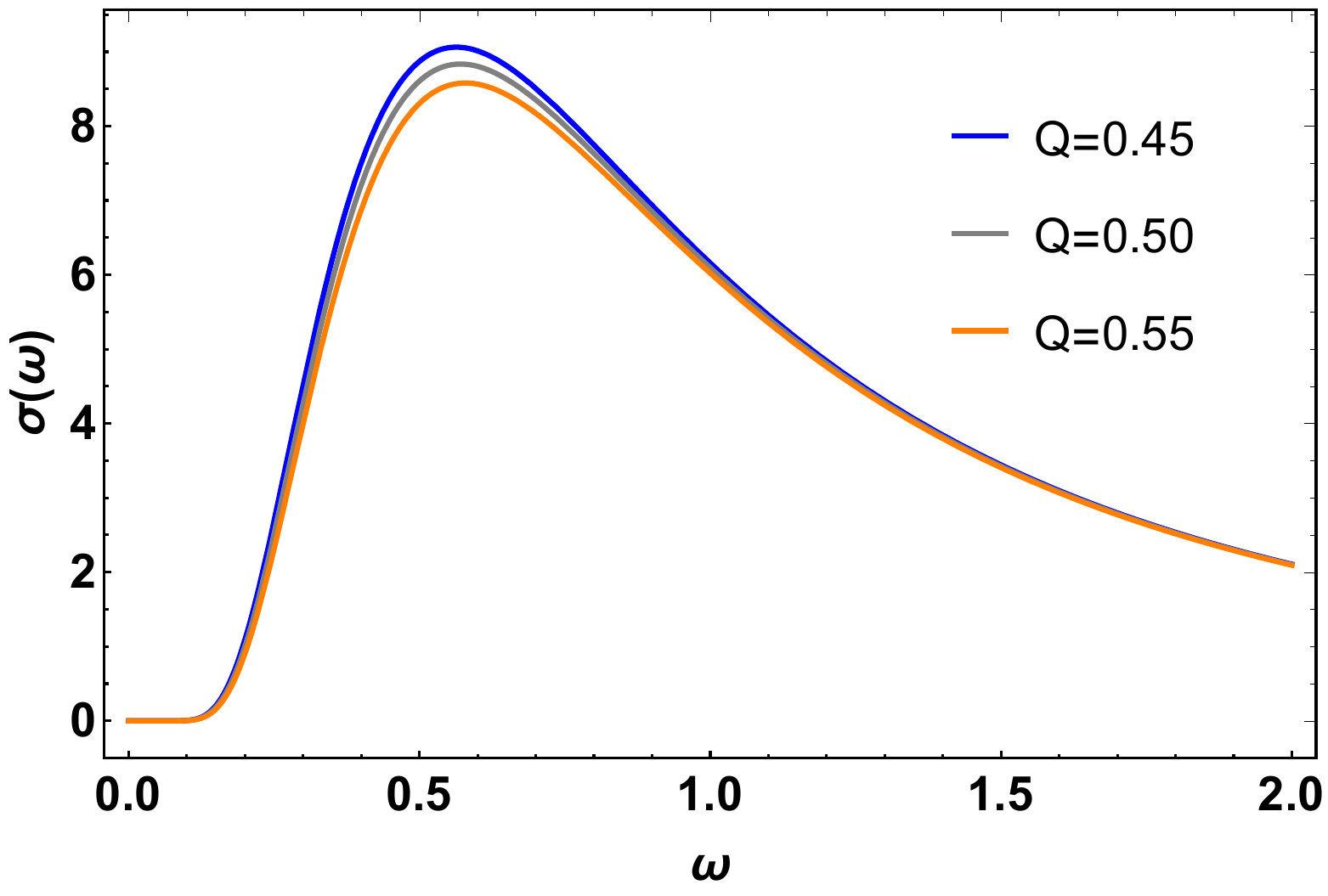}\\
(a)\hspace{6.7cm}(b)\\
\includegraphics[height=5cm]{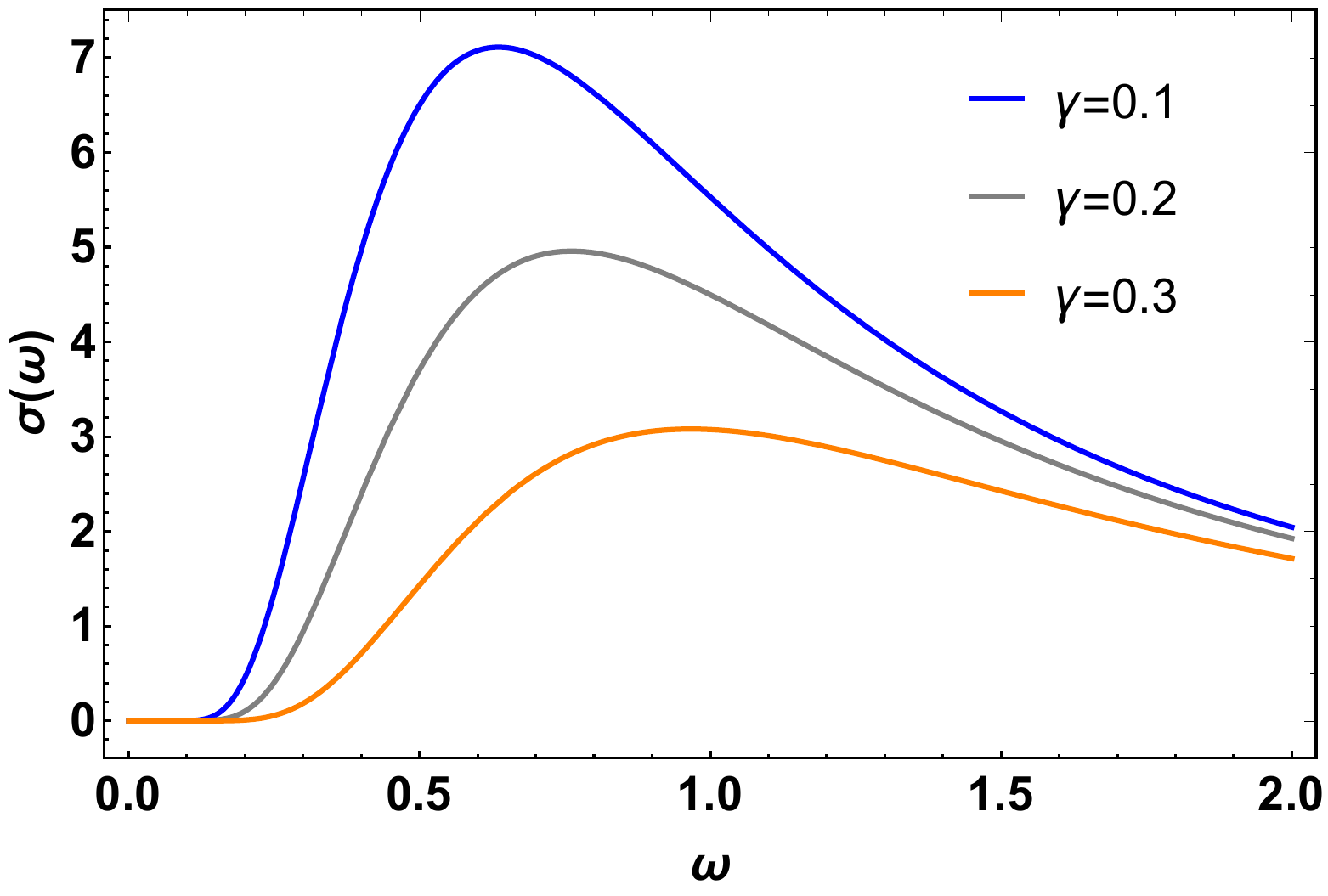} 
\includegraphics[height=5cm]{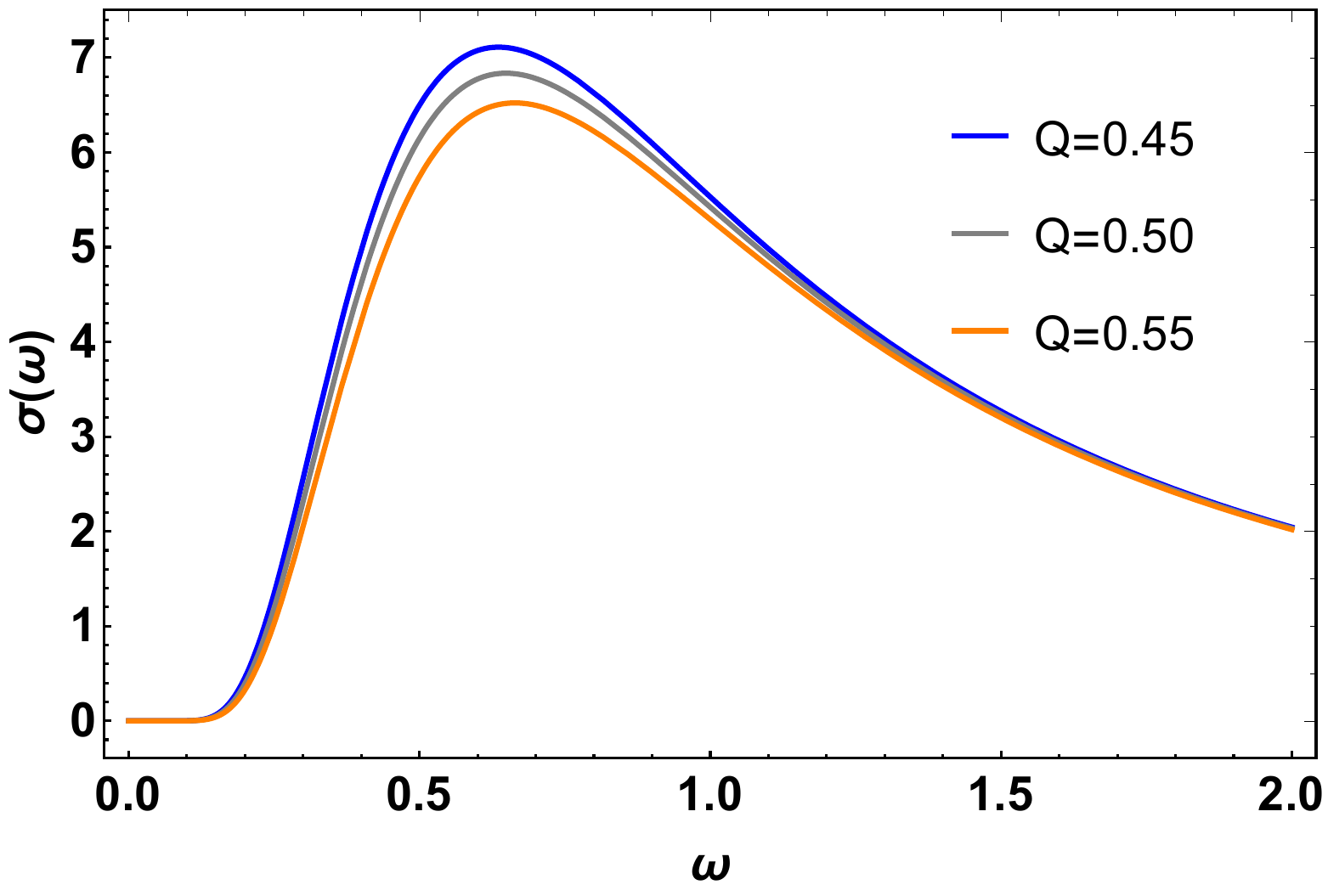}\\
(c)\hspace{6.7cm}(d)
\end{tabular}
\end{center}
\vspace{-0.5cm}
\caption{ (a) Absorption cross section for solution 1 with $l$ varying and $\lambda=0$ and $Q=0.45$. (b) Absorption cross section for solution 1 with $Q$ varying and $\lambda=0$ and $l=0.1$.
\label{fig4}}
\end{figure}

With the expressions of the greybody factor in hands, we can also calculate the absorption cross section through the following expression
\begin{align}
    \sigma(\omega)=\frac{\pi(2\lambda+1)}{\omega^2}\ T(\omega)
\end{align}
where $T(\omega)$ is the greybody factor. The panels (a) and (c) of Fig.\ref{fig4} depicts the absorption cross section $\sigma_{\rm abs}$ as a function of the frequency $\omega$, for three value of the LV parameters $l$ and $\gamma$, e.g. $l$ and $\gamma=0.1,\,0.2,\,0.3$. On the other hand, The panels (b) and (d) of Fig.\ref{fig4} depicts the absorption cross section $\sigma_{\rm abs}$ as a function of the frequency $\omega$ for three value of the electric charge $Q$, e.g. $Q=0.45,\,0.50,\,0.55$.

\begin{figure}[ht!]
\begin{center}
\begin{tabular}{ccc}
\includegraphics[height=5cm]{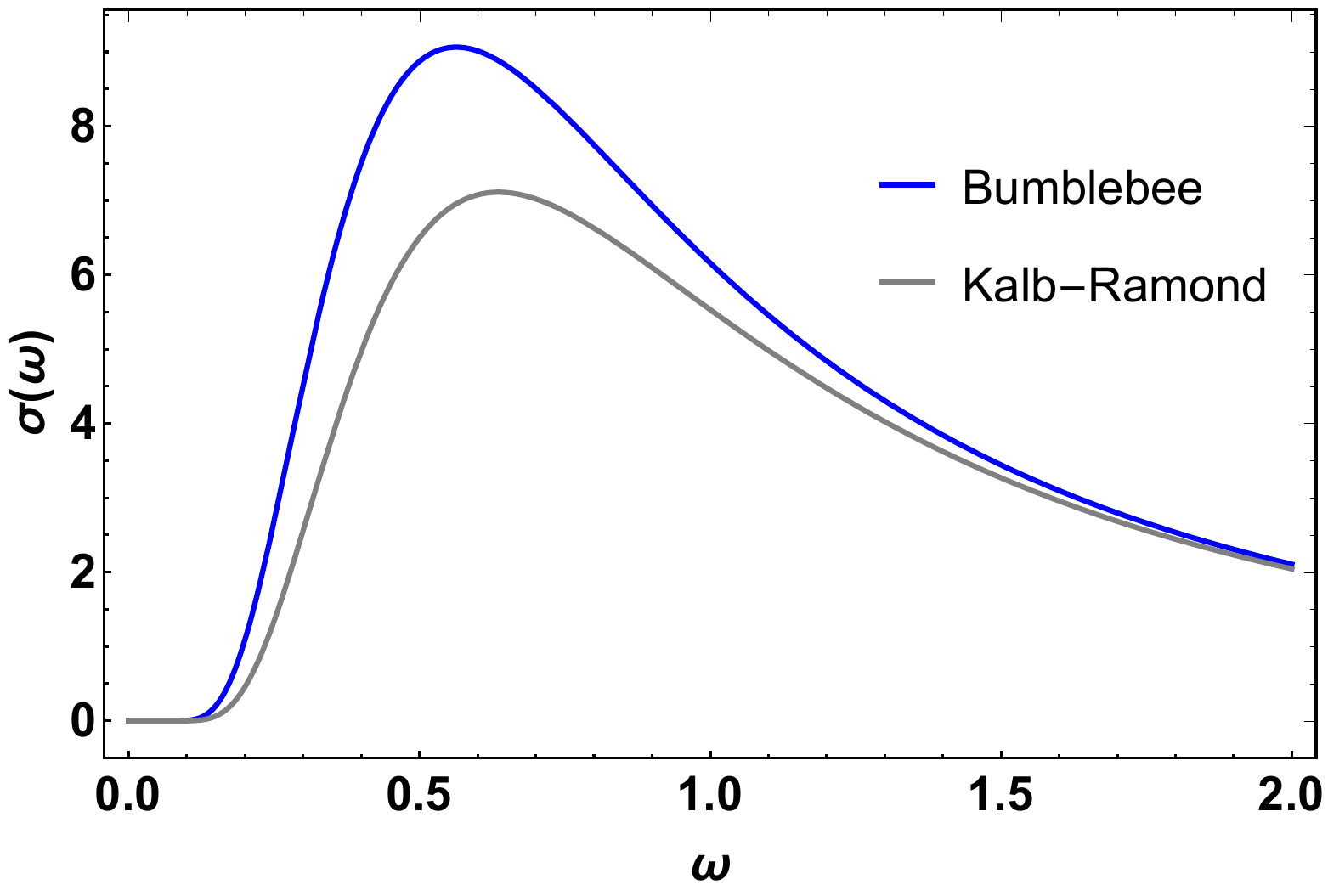} 
\end{tabular}
\end{center}
\vspace{-0.5cm}
\caption{ Comparison between the absorption cross section for bumblebee model with $\lambda=1$ and $l=0.1$ and the greybody factor for Kalb-Ramond model for $\lambda=1$ and $\gamma=0.1$. 
\label{fig5}}
\end{figure}

We note the trend that the LV parameters $l$ and $\gamma$ diminish the absorption cross section they increase. In the panel (a) of Fig.\ref{fig4}, we see easily that the peak value of $\sigma_{\rm abs}$ decreases considerable as one goes from smaller to larger $\gamma$. There is still a tiny shift of the onset/maximum toward higher $\omega$, what indicates that a larger frequency is necessary to reach a commensurate transmission when LV effects are stronger. In this sense, we interpret naturally that the parameter $\gamma$ makes the propagation of scalar particles more reflective. This result is equivalent to say that the LV parameter reduces greybody factors as we discussed previously. This alters both the effective radial potential and the dispersion relation of the scalar mode. Thereupon, a smaller fraction of the incoming scalar flux reaches the horizon, reducing $\sigma_{\rm abs}(\omega)$. Another noteworthy fact we should mention is that at larger $\omega$, LV effects is able to introduce additional dependence on frequency through the flux normalization and modified dispersion, so the absorbed flux per unit incident flux may decrease with frequency. Following this line of reasoning, a similar trend is also observed when varying the electric charge $Q$ (see panels (c) and (d) of Fig. \ref{fig4}). Furthermore, in Fig. \ref{fig5}, we see that the absorption cross-section for the Bumblebee model is smaller than for the Kalb-Ramond model.

It is important to point out that the analytical results obtained in this work are based on the low-frequency and weak-coupling approximations, allowing us to treat the wave equation solution in closed form. The partial-wave method and the analytic expression for the transmission probability are  valid when the incident scalar wave has a wavelength much larger than the characteristic black hole size, that is, $\omega M \ll 1$, where $M$ is the black hole mass and $\omega$ the wave frequency. In this regime, the potential barrier varies slowly on the scale of the wavelength, allowing one to truncate the expansion of the radial equation to the leading-order terms in $\omega$. Under these conditions, the analytic sech$^2$ estimate for the greybody factor provides an excellent approximation, as confirmed by direct numerical integration of the wave equation. For intermediate frequencies ($\omega M \sim 0.3$), small deviations appear, typically of order $5$–$10\%$ depending on the model parameters, while for $\omega M > 1$ higher partial waves dominate and the analytical formula ceases to be reliable.

Furthermore, the accuracy of the results depends on the smallness of the LV parameters $l$ (for the bumblebee model) and $\gamma$ (for the Kalb--Ramond case), which were treated perturbatively with $l,\gamma \ll 1$. This assumption is fundamental since it ensures that the spacetime remains asymptotically flat and that the corrections to the effective potential remain subleading with respect to the contributions from general relativity. Although larger values of these parameters may be considered for illustrative purposes, such cases should be interpreted only qualitatively and not as physically realistic. The perturbative expansion together with the low-frequency regime guarantees the mathematical consistency of the analysis.

These results are particularly relevant in the context of high-precision black-hole observations. Experiments such as the Event Horizon Telescope (EHT) and gravitational-wave detectors, including LIGO, Virgo, and the future LISA mission, provide indirect probes of the near-horizon structure of compact objects. Small deviations in the energy flux, quasinormal-mode spectra, or emission efficiency could signal new physics beyond general relativity. In this sense, the comparative framework developed here, which relates symmetry breaking, curvature deformations, and radiation emission, is able to offer a phenomenological setting in which LSV may be tested against astrophysical observations.

\section{Final remarks}\label{s6}

In this paper, we have delved into the impact of self-interacting vector and tensor fields with spontaneous Lorentz symmetry breaking on the properties of black holes with electric charge. In our analysis, one focused mainly on some relevant features, scattering, absorption, and greybody factors. Our main goal is to explore the physical implications of Lorentz symmetry breaking in the gravitational context, mainly the physics of black holes.

Initially, we employed the partial waves method to examine the processes of scattering and absorption. For the first solution corresponding to a charged black hole in bumblebee gravity, one notices that the scattering cross-section increases as the LV parameter concerning the bumblebee field increases, whereas it decreases as the electrical charge increases. A similar behavior happens in the absorption. Likewise, when considering the second solution corresponding to a charged black hole in KR gravity, we observe that the scattering cross section decreases as the LV parameter concerning the KR field increases. In contrast, it decreases as the electrical charge increases. Additionally, one has studied the greybody factor of massless spin 0 scalar fields by showing the influence of LV parameters $l$ and $\gamma$, and electrical charge $Q$. For both solutions, it is observed that the GF increases as both the LV parameter and the electric charge $Q$ increase. 

It is important to emphasize that, in the analysis performed here, we employ LV parameters values usually ranging from $\gamma,\ell = 0,1$ to $0,3$ for the purpose of visualizing LSV effects on scattering and absorption observables. Although these larger values for LV parameters exceed current empirical bounds, they are useful for visualizing the qualitative trends and scaling behavior of the cross sections and greybody factors. For quantitative comparison with observational data, only the lower end of this range remains viable. This ensures that all conclusions drawn here remain physically consistent within the regime where the LV deformation acts as a small perturbation to the underlying Reissner–Nordström geometry. Furthermore, the validity of the approximations low-frequency limit, small LV parameters, and asymptotic flatness guarantees that the analytic methods used capture the leading-order physical behavior of scalar fields in Lorentz-violating black hole backgrounds. For higher frequencies or stronger LSV, our analysis must be extended numerically using the wave equation or other methods, such as higher-order WKB or continued-fraction methods. These extensions are left for future work, where the impact of LSV on high-frequency scattering, quasinormal spectra, and gravitational-wave emission can be examined in more detail.

From the phenomenological point of view, recent observations have been able to impose very stringent upper bounds on the magnitude of LSV in the gravitational sector. In this context, astrophysical observations such as the Event Horizon Telescope (EHT) measurements of the black hole shadows of M87* and Sgr~A*, the orbital precession of the S2 star around Sgr~A* measured by the GRAVITY collaboration, and the ringdown frequencies detected by the LIGO/Virgo/KAGRA interferometers all constrain deviations from the Kerr and Reissner–Nordström metrics at the level of $|\gamma|, |\ell| \lesssim 10^{-6}$. Solar-system tests such as the Gravity Probe B experiment and light deflection measurements further restrict these parameters to $|\gamma|, |\ell| \lesssim 10^{-12}$ in the weak-field regime.

Ultimately, it is worth mentioning that future work could extend this analysis in several directions. Firstly, we can perform a numerical integration of the scalar perturbation equation to obtain exact transmission coefficients and confirm the analytic trends derived here. Other direction involves the study of quasinormal modes and their damping behavior could offer complementary constraints on the LV parameters. We can still extend the present framework to rotating or cosmological backgrounds since it could reveal additional phenomenology associated with anisotropic symmetry breaking. Finally, we will investigate a suitable approach to investigate greybody factor of vector and tensor perturbations concerning the LV charged black hole solutions discussed in this present work.

\section*{Acknowledgments}
\hspace{0.5cm} The author F. M. Belchior would like to express gratitude to the Conselho Nacional de Desenvolvimento Cient\'{i}fico e Tecnol\'{o}gico CNPq for grant No. 161092/2021-7.


\begin{thebibliography}{99}

%\cite{Sotiriou:2008rp}
\bibitem{Sotiriou:2008rp}
T.~P.~Sotiriou and V.~Faraoni,
%``f(R) Theories Of Gravity,''
Rev. Mod. Phys. \textbf{82} (2010), 451-497
%doi:10.1103/RevModPhys.82.451
[arXiv:0805.1726 [gr-qc]].

%\cite{Clifton:2011jh}
\bibitem{Clifton:2011jh}
T.~Clifton, P.~G.~Ferreira, A.~Padilla and C.~Skordis,
%``Modified Gravity and Cosmology,''
Phys. Rept. \textbf{513} (2012), 1-189
%doi:10.1016/j.physrep.2012.01.001
[arXiv:1106.2476 [astro-ph.CO]].

%\cite{Akbar:2006mq}
\bibitem{Akbar:2006mq}
M.~Akbar and R.~G.~Cai,
%``Thermodynamic Behavior of Field Equations for f(R) Gravity,''
Phys. Lett. B \textbf{648} (2007), 243-248
%doi:10.1016/j.physletb.2007.03.005
[arXiv:gr-qc/0612089 [gr-qc]].


%\cite{Kanti:1995vq}
\bibitem{Kanti:1995vq}
P.~Kanti, N.~E.~Mavromatos, J.~Rizos, K.~Tamvakis and E.~Winstanley,
%``Dilatonic black holes in higher curvature string gravity,''
Phys. Rev. D \textbf{54}, 5049-5058 (1996)
%doi:10.1103/PhysRevD.54.5049
[arXiv:hep-th/9511071 [hep-th]].

%\cite{Nojiri:2005vv}
\bibitem{Nojiri:2005vv}
S.~Nojiri, S.~D.~Odintsov and M.~Sasaki,
%``Gauss-Bonnet dark energy,''
Phys. Rev. D \textbf{71}, 123509 (2005)
%doi:10.1103/PhysRevD.71.123509
[arXiv:hep-th/0504052 [hep-th]].

%\cite{Belchior:2024bcn}
\bibitem{Belchior:2024bcn}
F.~M.~Belchior, R.~V.~Maluf, A.~Y.~Petrov and P.~J.~Porf\'\i{}rio,
%``Geometrical deformation of brane matter field within f(R,~Q,~P) gravity,''
Eur. Phys. J. C \textbf{84}, no.12, 1307 (2024)
%doi:10.1140/epjc/s10052-024-13684-8
[arXiv:2410.00743 [gr-qc]].


%\cite{Hawking:1975vcx}
\bibitem{Hawking:1975vcx}
S.~W.~Hawking,
%``Particle Creation by Black Holes,''
Commun. Math. Phys. \textbf{43}, 199-220 (1975)
[erratum: Commun. Math. Phys. \textbf{46}, 206 (1976)]
%doi:10.1007/BF02345020

%\cite{Gibbons:1977mu}
\bibitem{Gibbons:1977mu}
G.~W.~Gibbons and S.~W.~Hawking,
%``Cosmological Event Horizons, Thermodynamics, and Particle Creation,''
Phys. Rev. D \textbf{15}, 2738-2751 (1977)
%doi:10.1103/PhysRevD.15.2738


%\cite{Bekenstein:1973ur}
\bibitem{Bekenstein:1973ur}
J.~D.~Bekenstein,
%``Black holes and entropy,''
Phys. Rev. D \textbf{7}, 2333-2346 (1973)
%doi:10.1103/PhysRevD.7.2333


%\cite{LIGOScientific:2016aoc}
\bibitem{LIGOScientific:2016aoc}
B.~P.~Abbott \textit{et al.} [LIGO Scientific and Virgo],
%``Observation of Gravitational Waves from a Binary Black Hole Merger,''
Phys. Rev. Lett. \textbf{116}, no.6, 061102 (2016)
%doi:10.1103/PhysRevLett.116.061102
[arXiv:1602.03837 [gr-qc]].

%\cite{LIGOScientific:2017ycc}
\bibitem{LIGOScientific:2017ycc}
B.~P.~Abbott \textit{et al.} [LIGO Scientific and Virgo],
%``GW170814: A Three-Detector Observation of Gravitational Waves from a Binary Black Hole Coalescence,''
Phys. Rev. Lett. \textbf{119}, no.14, 141101 (2017)
%doi:10.1103/PhysRevLett.119.141101
[arXiv:1709.09660 [gr-qc]].

%\cite{LIGOScientific:2020zkf}
\bibitem{LIGOScientific:2020zkf}
R.~Abbott \textit{et al.} [LIGO Scientific and Virgo],
%``GW190814: Gravitational Waves from the Coalescence of a 23 Solar Mass Black Hole with a 2.6 Solar Mass Compact Object,''
Astrophys. J. Lett. \textbf{896}, no.2, L44 (2020)
%doi:10.3847/2041-8213/ab960f
[arXiv:2006.12611 [astro-ph.HE]].

%\cite{LIGOScientific:2018mvr}
\bibitem{LIGOScientific:2018mvr}
B.~P.~Abbott \textit{et al.} [LIGO Scientific and Virgo],
%``GWTC-1: A Gravitational-Wave Transient Catalog of Compact Binary Mergers Observed by LIGO and Virgo during the First and Second Observing Runs,''
Phys. Rev. X \textbf{9}, no.3, 031040 (2019)
%doi:10.1103/PhysRevX.9.031040
[arXiv:1811.12907 [astro-ph.HE]].

%\cite{LIGOScientific:2016dsl}
\bibitem{LIGOScientific:2016dsl}
B.~P.~Abbott \textit{et al.} [LIGO Scientific and Virgo],
%``Binary Black Hole Mergers in the first Advanced LIGO Observing Run,''
Phys. Rev. X \textbf{6}, no.4, 041015 (2016)
[erratum: Phys. Rev. X \textbf{8}, no.3, 039903 (2018)]
%doi:10.1103/PhysRevX.6.041015
[arXiv:1606.04856 [gr-qc]].

%\cite{VIRGO:2014yos}
\bibitem{VIRGO:2014yos}
F.~Acernese \textit{et al.} [VIRGO],
%``Advanced Virgo: a second-generation interferometric gravitational wave detector,''
Class. Quant. Grav. \textbf{32}, no.2, 024001 (2015)
%doi:10.1088/0264-9381/32/2/024001
[arXiv:1408.3978 [gr-qc]].

%\cite{Bluhm:2004ep}
\bibitem{Bluhm:2004ep}
R.~Bluhm and V.~A.~Kostelecky,
%``Spontaneous Lorentz violation, Nambu-Goldstone modes, and gravity,''
Phys. Rev. D \textbf{71}, 065008 (2005)
%doi:10.1103/PhysRevD.71.065008
[arXiv:hep-th/0412320 [hep-th]].

%\cite{Bailey:2006fd}
\bibitem{Bailey:2006fd}
Q.~G.~Bailey and V.~A.~Kostelecky,
%``Signals for Lorentz violation in post-Newtonian gravity,''
Phys. Rev. D \textbf{74}, 045001 (2006)
%doi:10.1103/PhysRevD.74.045001
[arXiv:gr-qc/0603030 [gr-qc]].


%\cite{Bluhm:2007bd}
\bibitem{Bluhm:2007bd}
R.~Bluhm, S.~H.~Fung and V.~A.~Kostelecky,
%``Spontaneous Lorentz and Diffeomorphism Violation, Massive Modes, and Gravity,''
Phys. Rev. D \textbf{77}, 065020 (2008)
%doi:10.1103/PhysRevD.77.065020
[arXiv:0712.4119 [hep-th]].

%\cite{Kostelecky:2010ze}
\bibitem{Kostelecky:2010ze}
A.~V.~Kostelecky and J.~D.~Tasson,
%``Matter-gravity couplings and Lorentz violation,''
Phys. Rev. D \textbf{83}, 016013 (2011)
%doi:10.1103/PhysRevD.83.016013
[arXiv:1006.4106 [gr-qc]].



\bibitem{Maluf:2020kgf}
R.~V.~Maluf and J.~C.~S.~Neves,
%``Black holes with a cosmological constant in bumblebee gravity,''
Phys. Rev. D \textbf{103} (2021) no.4, 044002
%doi:10.1103/PhysRevD.103.044002
[arXiv:2011.12841 [gr-qc]].

\bibitem{Casana:2017jkc}
R.~Casana, A.~Cavalcante, F.~P.~Poulis and E.~B.~Santos,
%``Exact Schwarzschild-like solution in a bumblebee gravity model,''
Phys. Rev. D \textbf{97} (2018) no.10, 104001
%doi:10.1103/PhysRevD.97.104001
[arXiv:1711.02273 [gr-qc]].


%\cite{Gullu:2020qzu}
\bibitem{Gullu:2020qzu}
\.I.~G\"ull\"u and A.~\"Ovg\"un,
%``Schwarzschild-like black hole with a topological defect in bumblebee gravity,''
Annals Phys. \textbf{436}, 168721 (2022)
%doi:10.1016/j.aop.2021.168721
[arXiv:2012.02611 [gr-qc]].

%\cite{Liu:2024axg}
\bibitem{Liu:2024axg}
J.~Z.~Liu, W.~D.~Guo, S.~W.~Wei and Y.~X.~Liu,
%``Charged spherically symmetric and slowly rotating charged black hole solutions in bumblebee gravity,''
Eur. Phys. J. C \textbf{85}, no.2, 145 (2025)
%doi:10.1140/epjc/s10052-025-13859-x
[arXiv:2407.08396 [gr-qc]].


\bibitem{Altschul:2009ae}
B.~Altschul, Q.~G.~Bailey and V.~A.~Kostelecky,
%``Lorentz violation with an antisymmetric tensor,''
Phys. Rev. D \textbf{81} (2010), 065028
%doi:10.1103/PhysRevD.81.065028
[arXiv:0912.4852 [gr-qc]].


\bibitem{Maluf:2018jwc}
R.~V.~Maluf, A.~A.~Ara\'ujo Filho, W.~T.~Cruz and C.~A.~S.~Almeida,
%``Antisymmetric tensor propagator with spontaneous Lorentz violation,''
EPL \textbf{124} (2018) no.6, 61001
%doi:10.1209/0295-5075/124/61001
[arXiv:1810.04003 [hep-th]].


\bibitem{Aashish:2019ykb}
S.~Aashish and S.~Panda,
%``Quantum aspects of antisymmetric tensor field with spontaneous Lorentz violation,''
Phys. Rev. D \textbf{100} (2019) no.6, 065010
%doi:10.1103/PhysRevD.100.065010
[arXiv:1903.11364 [gr-qc]].


\bibitem{Lessa:2019bgi}
L.~A.~Lessa, J.~E.~G.~Silva, R.~V.~Maluf and C.~A.~S.~Almeida,
%``Modified black hole solution with a background Kalb\textendash{}Ramond field,''
Eur. Phys. J. C \textbf{80} (2020) no.4, 335
%doi:10.1140/epjc/s10052-020-7902-1
[arXiv:1911.10296 [gr-qc]].

%\cite{Yang:2023wtu}
\bibitem{Yang:2023wtu}
K.~Yang, Y.~Z.~Chen, Z.~Q.~Duan and J.~Y.~Zhao,
%``Static and spherically symmetric black holes in gravity with a background Kalb-Ramond field,''
Phys. Rev. D \textbf{108} (2023) no.12, 124004
%doi:10.1103/PhysRevD.108.124004
[arXiv:2308.06613 [gr-qc]].

%\cite{Belchior:2025xam}
\bibitem{Belchior:2025xam}
F.~M.~Belchior, R.~V.~Maluf, A.~Y.~Petrov and P.~J.~Porf\'\i{}rio,
%``Global monopole in a Ricci-coupled Kalb-Ramond bumblebee gravity,''
[arXiv:2502.17267 [gr-qc]].

%\cite{Duan:2023gng}
\bibitem{Duan:2023gng}
Z.~Q.~Duan, J.~Y.~Zhao and K.~Yang,
%``Electrically charged black holes in gravity with a background Kalb\textendash{}Ramond field,''
Eur. Phys. J. C \textbf{84} (2024) no.8, 798
%doi:10.1140/epjc/s10052-024-13188-5
[arXiv:2310.13555 [gr-qc]].


%\cite{Sanchez:1976xm}
\bibitem{Sanchez:1976xm}
N.~G.~Sanchez,
%``The Wave Scattering Theory and the Absorption Problem for a Black Hole,''
Phys. Rev. D \textbf{16} (1977), 937-945.
%doi:10.1103/PhysRevD.16.937

%\cite{Crispino:2009ki}
\bibitem{Crispino:2009ki}
L.~C.~B.~Crispino, S.~R.~Dolan and E.~S.~Oliveira,
%``Scattering of massless scalar waves by Reissner-Nordstr\"om black holes,''
Phys. Rev. D \textbf{79}, 064022 (2009)
%doi:10.1103/PhysRevD.79.064022
[arXiv:0904.0999 [gr-qc]].


\bibitem{Anacleto:2020zhp}
M.~A.~Anacleto, F.~A.~Brito, J.~A.~V.~Campos and E.~Passos,
%``Absorption and scattering by a self-dual black hole,''
Gen. Rel. Grav. \textbf{52}, no.10, 100 (2020)
%doi:10.1007/s10714-020-02756-1
[arXiv:2002.12090 [hep-th]].

%\cite{Anacleto:2019tdj}
\bibitem{Anacleto:2019tdj}
M.~A.~Anacleto, F.~A.~Brito, J.~A.~V.~Campos and E.~Passos,
%``Absorption and scattering of a noncommutative black hole,''
Phys. Lett. B \textbf{803} (2020), 135334
%doi:10.1016/j.physletb.2020.135334
[arXiv:1907.13107 [hep-th]].

%\cite{Pitelli:2017bgx}
\bibitem{Pitelli:2017bgx}
J.~P.~M.~Pitelli, V.~S.~Barroso and M.~Richartz,
%``Scattering Cross Section and Stability of Global Monopoles,''
Phys. Rev. D \textbf{96} (2017) no.10, 105021
%doi:10.1103/PhysRevD.96.105021
[arXiv:1711.03526 [gr-qc]].

%\cite{Anacleto:2017kmg}
\bibitem{Anacleto:2017kmg}
M.~A.~Anacleto, F.~A.~Brito, S.~J.~S.~Ferreira and E.~Passos,
%``Absorption and scattering of a black hole with a global monopole in f(R) gravity,''
Phys. Lett. B \textbf{788} (2019), 231-237
%doi:10.1016/j.physletb.2018.11.020
[arXiv:1701.08147 [hep-th]].

%\cite{Anacleto:2022shk}
\bibitem{Anacleto:2022shk}
M.~A.~Anacleto, F.~A.~Brito, J.~A.~V.~Campos and E.~Passos,
%``Absorption, scattering and shadow by a noncommutative black hole with global monopole,''
Eur. Phys. J. C \textbf{83} (2023) no.4, 298
%doi:10.1140/epjc/s10052-023-11484-0
[arXiv:2212.13973 [hep-th]].

%\cite{Boonserm:2008zg}
\bibitem{Boonserm:2008zg}
P.~Boonserm and M.~Visser,
%``Bounding the greybody factors for Schwarzschild black holes,''
Phys. Rev. D \textbf{78}, 101502 (2008)
%doi:10.1103/PhysRevD.78.101502
[arXiv:0806.2209 [gr-qc]].

%\cite{Okyay:2021nnh}
\bibitem{Okyay:2021nnh}
M.~Okyay and A.~\"Ovg\"un,
%``Nonlinear electrodynamics effects on the black hole shadow, deflection angle, quasinormal modes and greybody factors,''
JCAP \textbf{01}, no.01, 009 (2022)
%doi:10.1088/1475-7516/2022/01/009
[arXiv:2108.07766 [gr-qc]].

%\cite{Guo:2023nkd}
\bibitem{Guo:2023nkd}
W.~D.~Guo, Q.~Tan and Y.~X.~Liu,
%``Quasinormal modes and greybody factor of a Lorentz-violating black hole,''
JCAP \textbf{07} (2024), 008
%doi:10.1088/1475-7516/2024/07/008
[arXiv:2312.16605 [gr-qc]].

%\cite{Wu:2024ldo}
\bibitem{Wu:2024ldo}
L.~B.~Wu, R.~G.~Cai and L.~Xie,
%``Stability of the greybody factor of Hayward black holes,''
Phys. Rev. D \textbf{111}, no.4, 044066 (2025)
%doi:10.1103/PhysRevD.111.044066
[arXiv:2411.07734 [gr-qc]].


%\cite{Sucu:2025lqa}
\bibitem{Sucu:2025lqa}
E.~Sucu and {\.I}.~Sakall,
%``Exploring Lorentz-violating effects of Kalb-Ramond field on charged black hole thermodynamics and photon dynamics,''
Phys. Rev. D \textbf{111}, no.6, 064049 (2025)
%doi:10.1103/PhysRevD.111.064049

%\cite{Li:2024xyu}
\bibitem{Li:2024xyu}
Q.~Li, Q.~Wang and J.~Jia,
%``Absorption and scattering of charged scalar waves by charged Horndeski black hole,''
Phys. Rev. D \textbf{111}, no.2, 024059 (2025)
%doi:10.1103/PhysRevD.111.024059
[arXiv:2411.02987 [gr-qc]].

%\cite{Campos:2023zmg}
\bibitem{Campos:2023zmg}
J.~A.~V.~Campos, M.~A.~Anacleto, F.~A.~Brito and E.~Passos,
%``Absorption, scattering, quasinormal modes and shadow by canonical acoustic black holes in Lorentz-violating background,''
Gen. Rel. Grav. \textbf{56}, no.6, 74 (2024)
%doi:10.1007/s10714-024-03263-3
[arXiv:2312.14258 [gr-qc]].

%\cite{al-Badawi:2024pdx}
\bibitem{al-Badawi:2024pdx}
A.~al-Badawi, S.~Shaymatov and I.~Sakall{\i},
%``Geodesics structure and deflection angle of electrically charged black holes in gravity with a background Kalb{\textendash}Ramond field,''
Eur. Phys. J. C \textbf{84}, no.8, 825 (2024)
%doi:10.1140/epjc/s10052-024-13205-7
[arXiv:2408.09228 [gr-qc]].

%\cite{Richarte:2021fbi}
\bibitem{Richarte:2021fbi}
M.~G.~Richarte, {\'E}.~L.~Martins and J.~C.~Fabris,
%``Scattering and absorption of a scalar field impinging on a charged black hole in the Einstein-Maxwell-dilaton theory,''
Phys. Rev. D \textbf{105}, no.6, 064043 (2022)
%doi:10.1103/PhysRevD.105.064043
[arXiv:2111.01595 [gr-qc]].


\end{thebibliography}
\end{document}